\documentclass[aip,cha,floatfix,reprint,amssymb,amsmath]{revtex4-1}

\usepackage{graphicx,color}
\usepackage[normalem]{ulem}

\begin{document}

\title{Energy Distribution in Intrinsically Coupled Systems: The Spring Pendulum Paradigm}

\author{M. C. de Sousa}
\email[]{meirielenso@gmail.com}
\affiliation{Institute of Physics, University of S\~ao Paulo, 05508-090, S\~ao Paulo, S\~ao Paulo, Brazil}
\author{F. A. Marcus}
\email[]{albertus.marcus@gmail.com}
\thanks{Current affiliation: Department of Physics, Technological Institute of Aeronautics, 12228-900, S\~ao Jos\'e dos Campos, S\~ao Paulo, Brazil}
\affiliation{Department of Physics, Federal University of Paran\'a, C.P. 19044, 81531-990, Curitiba, Paran\'a, Brazil}
\author{I. L. Caldas}
\affiliation{Institute of Physics, University of S\~ao Paulo, 05508-090, S\~ao Paulo, S\~ao Paulo, Brazil}
\author{R. L. Viana}
\affiliation{Department of Physics, Federal University of Paran\'a, C.P. 19044, 81531-990, Curitiba, Paran\'a, Brazil}

\date{\today}

\begin{abstract}
Intrinsically nonlinear coupled systems present different oscillating components that exchange energy among themselves. We present a new approach to deal with such energy exchanges and to investigate how it depends on the system control parameters. The method consists in writing the total energy of the system, and properly identifying the energy terms for each component and, especially, their coupling. To illustrate the proposed approach, we work with the bi-dimensional spring pendulum, which is a paradigm to study nonlinear coupled systems, and is used as a model for several systems. For the spring pendulum, we identify three energy components, resembling the spring and pendulum like motions, and the coupling between them. With these analytical expressions, we analyze the energy exchange for individual trajectories, and we also obtain global characteristics of the spring pendulum energy distribution by calculating spatial and time average energy components for a great number of trajectories (periodic, quasi-periodic and chaotic) throughout the phase space. Considering an energy term due to the nonlinear coupling, we identify regions in the parameter space that correspond to strong and weak coupling. The presented procedure can be applied to nonlinear coupled systems to reveal how the coupling mediates internal energy exchanges, and how the energy distribution varies according to the system parameters.

\end{abstract}

\maketitle

\section{Introduction}
\label{sec:Introduction}

Nonlinear coupled systems with an arbitrary number of interacting subsystems are present in many areas, from physics and engineering to biology and social sciences. Examples of coupled systems include wave coupling in plasma physics \cite{Sagdeev1969,Ritz1986,Ritz1988,Horton2012}, coupled lasers \cite{Wiesenfeld1990,Kozyreff2000,Zamora-Munt2010}, biological oscillator networks \cite{Winfree1980,Kuramoto1984,Strogatz1993,Bressloff1997,Newman2010}, neural networks \cite{Abbott1993,Collins1995,Joya2002,Wang2010}, and genetic networks \cite{Bolouri2002,DeJong2002,Ren2008}.

Coupled systems usually present properties that are not found in the individual subsystems. The new properties of coupled systems depend on the coupling and the energy exchanges among the subsystems. In the literature, we find studies about energy exchanges in nonlinear coupled systems \cite{Ford1961,Jackson1963,Ritz1986,Ritz1988,Gendelman2001,Vakakis2004,Quinn2008,Kovaleva2010,Sigalov2012}. However, most of these studies focus on analytical approximations for weakly coupled systems and the energy exchanges that occur when the subsystems are in resonance. In this context, it is a challenge to investigate, for both weak and strong coupling, as well as for high energies and large amplitude oscillations, the energy distribution among the components of nonlinear coupled systems.

A very efficient mechanism of energy exchange is the parametric mechanism \cite{Gendelman2001}. In particular, the spring pendulum, also known as elastic or extensible pendulum, with two degrees of freedom is an autoparametric system that represents a paradigm for the study of nonlinear coupled systems. The spring pendulum presents many interesting dynamical features, such as energy exchange in the parametric resonance condition \cite{Vitt1933,Kane_JAM1968,Tselman_JAMM1970,Rusbridge_AmJP1980,Breitenberger_JMP1981,Lai_AmJP1984}, and an order-chaos-order transition as the system parameters increase \cite{Yepez_PLA1990,Cuerno_AmJP92,Gonzalez_EJP94,Weele_PhysA1996}. Furthermore, the spring pendulum is relevant due to its qualitative representation of many nonlinear coupled systems of great physical interest.

Among these representations, some examples are the orbits of celestial bodies \cite{Contopoulos_AJ1963,Hitzl_CM1975}, such as satellites (both natural and artificial) and asteroids \cite{Hori_ASJ1966,Broucke_CM1973}, the classical analogue for the vibrational modes of triatomic molecules producing the Fermi resonance in the infrared and Raman spectra \cite{Ramaneffekt1931,Amat_JMS1965,Jacob_JPB1978}, the interaction between light waves in a nonlinear medium \cite{Armstrong_PR1962}, and wave coupling in plasma physics \cite{Sagdeev1969,Horton2012}. In mechanical engineering, different types of pendulum are widely used \cite{Sigalov2012,Orosco2016,Orzechowski2015,Tusset2016,Nishimura2016,Rocha2017}. In particular, the spring pendulum is used both as a component of mechanical systems \cite{Anh2007}, as well as a model whose equations of motion describe the behavior of several mechanical devices \cite{Holmes2006,Wang2011,Castillo-Rivera2017}.

In this paper, we present a new approach to investigate energy exchanges in nonlinear coupled systems. To illustrate the method, we use the spring pendulum because of the richness of its complex behavior. We analyze the coupling in spring pendulums and how it mediates energy exchanges between the spring-mass and pendular like motions. To do so, we consider the total energy distributed among the two subsystems, and we identify an energy term due to the nonlinear coupling. We describe the system using coordinates that relate directly to the spring and pendulum like motions, and we write the Hamiltonian as a sum of three terms, spring-mass, pendulum and coupling, resembling, respectively, the energy associated with the spring and pendulum motions and their coupling. Within this analysis, we find how the energy is distributed among the considered three energy terms, and how the energy distribution varies according to the total energy and a control parameter that represents the ratio of the simple pendulum and spring-mass frequencies.

Considering an energy term due to the nonlinear coupling, we identify a transition from strong to weak coupling as we increase the total energy. When the coupling in the system is strong, the spring-mass and the pendulum move as a unique new system and, most of the time, it is difficult to distinguish the two kinds of movement. For weak coupling, the spring-mass and the pendulum slightly interact with each other. In this case, we can identify the spring and pendulum individual motions for certain periods of time.

It is important to notice that the approach we propose can be applied to other nonlinear coupled systems to investigate the coupling and the energy distribution among the oscillating components. The method we present is valid for weak and strong coupling, low and high values of energy and oscillation amplitude, as well as all kinds of trajectories the coupled system may present (resonant islands, invariant tori and chaotic trajectories).

The mathematical description of the spring pendulum is presented in Section \ref{sec:DescripSpringPendulum}, where we introduce coordinates that relate directly to the spring and pendulum like motions. In Section \ref{sec:EnergyDistribution}, we discuss the coupled evolution of the spring-mass and pendulum subsystems. We distribute the total energy among three energy terms, spring, pendulum and coupling, and we justify our definitions for each energy term. In Section \ref{sec:ResultsDiscussion}, we investigate the energy distribution for different trajectories and parameters of the system. Using our definition for the energy terms, we calculate the average spring, pendulum and coupling energy terms and we analyze how the energy distribution varies according to the total energy and the control parameter. We obtain a scaling law for the coupling energy term, and we identify regions of strong and weak coupling in the parameter space of the system. Finally, we draw our conclusions in Section \ref{sec:Conclusions}.

\section{The Spring Pendulum}
\label{sec:DescripSpringPendulum}

An important issue concerning coupled systems is the energy distribution among the components due to a nonlinear coupling, and how this energy distribution varies according to some control parameter. We propose a new approach to investigate the energy distribution in coupled systems, and how the coupling mediates energy exchanges among the system components.
\begin{figure}[!tb]
	\centering
	\includegraphics[width=0.5\linewidth]{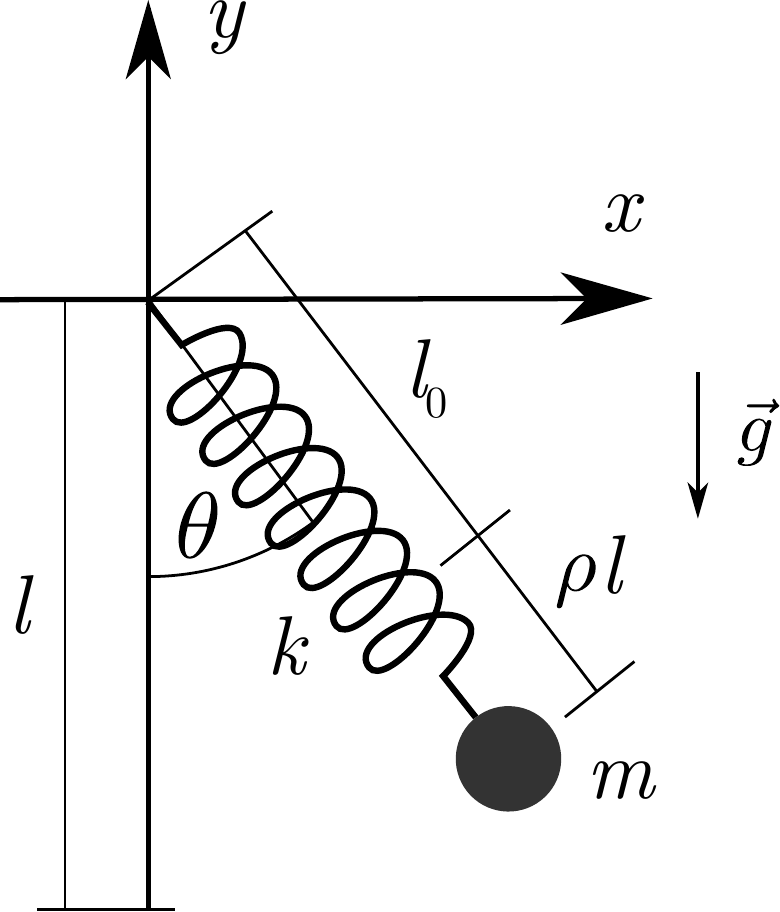}
	\caption{Schematic representation of a spring pendulum.}
	\label{fig:SpringPend}
\end{figure}

To present our approach, we consider a paradigm for the study of coupled systems with nonlinear characteristics: the spring pendulum with two degrees of freedom. The spring pendulum is composed of a mass $m$ attached to the free extremity of a massless spring with stiffness constant $k$, and length $l_0$ in the absence of forces. The other extremity of the spring is fixed at the center of the Cartesian coordinate system $(x,y)=(0,0)$ as shown in Figure \ref{fig:SpringPend}. The system moves only in the vertical plane, and its stable equilibrium position corresponds to $(x,y)=(0,-l)$, where $l=l_0+mg/k$, and $g$ is the acceleration of gravity. The Hamiltonian of this system in Cartesian coordinates is given by
\begin{equation}
	\label{eq:Hxy}
	\mathcal{E_T} = \mathcal{H} = \frac{p_x^2 + p_y^2}{2m} + mgy + \frac{k}{2}(\sqrt{x^2 + y^2} - l_0)^2 ,
\end{equation}
where $\mathcal{E_T}$ is the total energy.

To better understand how the total energy is distributed in the spring pendulum, we describe the system using dimensionless coordinates that relate directly to the spring-mass and pendulum motions:
\begin{gather} \label{eq:def_rho_theta}
	\begin{aligned}
		\rho &= f - 1 + \frac{\sqrt{x^2 + y^2}}{l} , \\*
		\theta &= \arctan \frac{x}{-y} .
	\end{aligned}
\end{gather}
The dimensionless momenta canonically conjugated to $\rho$ and $\theta$ are
\begin{gather} \label{eq:def_prho_ptheta}
	\begin{aligned}
		p_{\rho} &= \frac{d \rho}{dt} , \\*
		p_{\theta} &= \frac{d \theta}{dt} (\rho+1-f)^2 ,
	\end{aligned}
\end{gather}
with $t = \tau \sqrt{k/m}$ the dimensionless time variable.

In the polar coordinate system defined by expressions (\ref{eq:def_rho_theta}), $\rho$ represents the spring extension or compression from $l_0$, and $\theta$ is the angle formed between the mass $m$ and the vertical axis pointing down, as shown in Figure \ref{fig:SpringPend}. We define the parameter $f$ as $f=mg/kl$, and from the stable equilibrium condition $l=l_0+mg/k$, we have $l_0/l=1-f$. It implies that $f$ must be in the interval $]0,1[$, since $l_0/l$ is a positive quantity.
\begin{figure*}[!tb]
	\centering
	\includegraphics{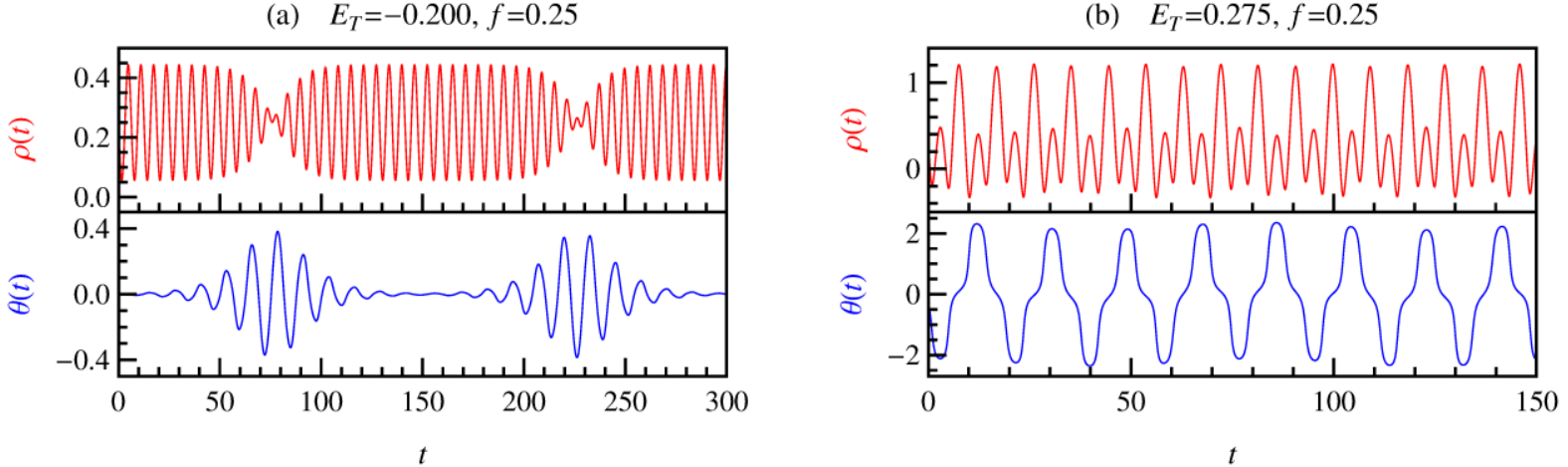}
	\caption{Time evolution of individual trajectories for $f=0.25$ and (a) $E_T=-0.200$, (b) $E_T=0.275$.}
	\label{fig:TimeSeries}
\end{figure*}

Using coordinates (\ref{eq:def_rho_theta}) and momenta (\ref{eq:def_prho_ptheta}), we rewrite expression (\ref{eq:Hxy}) as a dimensionless Hamiltonian $H=\mathcal{H}/kl^2$:
\begin{eqnarray} \label{eq:H_rho_theta}
	E_T = H & = & \frac{1}{2}\left[p_{\rho}^2 + \frac{p_{\theta}^2}{(\rho+1-f)^2}\right] + \frac{\rho^2}{2}      \nonumber  \\*
	        & - & (\rho+1-f)f\cos\theta \ ,
\end{eqnarray}
where $E_T$ is the dimensionless total energy.

The total energy is the only constant of motion of the non-integrable system with two degrees of freedom. Hamiltonian (\ref{eq:H_rho_theta}) written in the polar coordinates $(\rho, \theta)$ gives us a better view about the system and the two types of motion it presents: spring-mass and pendulum. However, its equations of motion are not very tractable for direct integration. To perform our analysis, we use a fourth-order symplectic integrator \cite{Forest1990} to solve the equations of motion in dimensionless Cartesian coordinates. The results in the $(\rho, \theta)$ coordinates are obtained through the canonical transformation (\ref{eq:def_rho_theta}). We point out that approximate analytical solutions for the equations of motion are only possible for restricted configurations of the spring pendulum, i.e. low total energy and small amplitude oscillations \cite{Vitt1933,Kane_JAM1968,Tselman_JAMM1970,Rusbridge_AmJP1980,Breitenberger_JMP1981,Lai_AmJP1984}. For all the other configurations, one should solve the equations of motion numerically.

In the spring pendulum, the spring-mass and pendular motions are coupled by the products of $\rho$ and $\cos \theta$, and $p_{\theta}$ by $(\rho+1-f)$, as can be seen from Hamiltonian (\ref{eq:H_rho_theta}) and the corresponding equations of motion. It is important to notice that this coupling is intrinsic i.e., the coupling arises from the configuration of the physical system. By replacing the fixed length rod of a simple pendulum with a spring, we create an intrinsically coupled system and, thus, the spring-mass and pendular motions exchange energy through the coupling. A system with such properties is known as autoparametric system \cite{Tondl_Book2000}. This intrinsic coupling is different from the usually considered coupling between two distinct oscillators.

\section{Energy distribution}
\label{sec:EnergyDistribution}

\subsection{Coupled Time Evolution}
\label{sec:CoupledEvolution}

The behavior of nonlinear coupled systems is governed by the coupling among the different subsystems. To understand the dynamics, it is necessary to know how the coupling acts on the system, and how it causes internal energy exchanges among the system components.

In the literature, most of the papers about spring pendulums study energy exchanges between the Cartesian coordinates $(x,y)$ for the parametric resonance \cite{Vitt1933,Kane_JAM1968,Tselman_JAMM1970,Rusbridge_AmJP1980,Breitenberger_JMP1981,Lai_AmJP1984}. For the method presented here, we work with coordinates that relate directly to the spring and pendulum movements, and we investigate the energy exchanges between the two subsystems for all kinds of trajectory.

In Figure \ref{fig:TimeSeries}.(a), we show the time evolution of the $(\rho, \theta)$ coordinates in the parametric resonance condition. For this trajectory, we observe that the spring energy is transferred to the pendular motion and back only in limited time intervals. When the spring transfers its energy to the pendulum, it remains almost still and only the pendular motion is appreciable. The opposite occurs when the pendulum transfers energy back to the spring. The pendulum moves just slightly, whereas the spring is compressed and stretched with great energy. This kind of behavior is known as autoparametric resonance \cite{Verhulst2002}. It occurs when the total energy is low, and the ratio of the simple pendulum and spring-mass frequencies is 1/2, which in our mathematical description of the system corresponds to $f=0.25$.

Figure \ref{fig:TimeSeries}.(b) shows the behavior of $(\rho, \theta)$ for a quasi-periodic trajectory that does not match the autoparametric resonance condition. In Figure \ref{fig:TimeSeries}.(b), both the spring and the pendulum are in constant motion and they exchange energy regularly. This scenario is much more common than the autoparametric resonance depicted in Figure \ref{fig:TimeSeries}.(a).

The autoparametric resonance is largely studied in the literature \cite{Vitt1933,Kane_JAM1968,Tselman_JAMM1970,Rusbridge_AmJP1980,Breitenberger_JMP1981,Lai_AmJP1984} because it allows one to obtain approximate analytical solutions to the spring pendulum equations of motion. However, the autoparametric resonance occurs only for very specific configurations of the system: low total energy, small amplitude oscillations and $f = 0.25$. For all the other configurations of the system, we have to solve the equations of motion numerically. In these cases, the energy exchanges between the spring-mass and pendulum motions have not been explored in the literature yet.

In this paper, we propose a new approach to analyze the coupling and the internal energy exchanges it causes. We define energy terms associated with each component of the system (i.e. spring-mass and pendulum), and an energy term due to the coupling. The analytical expressions we obtain for the energy terms are not restricted to the parametric resonance condition. They are valid for all values of total energy and $f$, and they can be used to describe both small and large oscillations, weak and strong coupling. With these energy terms, we are able to analyze the energy distribution for individual trajectories, and the average energy distribution for groups of trajectories.

\subsection{Spring, pendulum and coupling energy terms}
\label{sec:EnergyTerms}

In a simple pendulum, the length of the rod is fixed. In the spring pendulum, the fixed length rod is replaced by a spring whose length varies in time, coupling the spring and the pendulum motions. Since the spring-mass and the pendulum motions are nonlinearly coupled, we can regard the total energy terms in (\ref{eq:H_rho_theta}) as those resembling a spring-mass, a simple pendulum and the coupling between them.

Following this idea, we consider that the total energy $E_T$ of the spring pendulum is distributed among three distinct terms: spring-mass, pendulum and coupling. Accordingly, the spring energy term $E_S$ is the energy associated with a spring-mass system moving vertically under the action of gravity. The spring energy term represents the kinetic energy of the spring-mass, as well as its elastic and gravitational potential energy. It is a function of $(\rho, p_{\rho})$ only, and we write the spring energy term $E_S$ as
\begin{equation} \label{eq:ES}
	E_S = \frac{p_{\rho}^2 + \rho^2}{2} - (\rho+1-f)f.
\end{equation}

The pendulum energy term $E_P$ is a function of $(\theta, p_{\theta})$. It corresponds to the energy stored in a simple pendulum, in which the mass $m$ is suspended by a rod of fixed length $l$:
\begin{equation} \label{eq:EP}
	E_P = \frac{p_{\theta}^2}{2} - f\cos\theta.
\end{equation}
We choose $l$ as the fixed length of our simple pendulum model because this is the length of the extended spring in the stable equilibrium position of the spring pendulum described by Hamiltonian (\ref{eq:H_rho_theta}).

In the spring pendulum, the spring-mass and pendulum motions are nonlinearly coupled as can be seen in the second and fourth terms of Hamiltonian (\ref{eq:H_rho_theta}). The coupling in the spring pendulum is associated with the energy exchanges between the two kinds of movement described by the energy terms (\ref{eq:ES}) and (\ref{eq:EP}).

We define the coupling energy term $E_C$ for the spring pendulum as the amount of energy that arises from this nonlinear coupling:
\begin{eqnarray} \label{eq:EC}
	E_C & = & \frac{p_{\theta}^2}{2}\left[\frac{1}{(\rho+1-f)^2} - 1 \right] - (\rho-f)f\cos\theta      \nonumber  \\*
	    & + & (\rho+1-f)f.
\end{eqnarray}
As one would expect, the coupling energy term is a function of both the spring and pendulum coordinates: $\rho$, $\theta$, and $p_{\theta}$. Furthermore, using our definitions for the energy terms (\ref{eq:ES})-(\ref{eq:EC}), the total energy (\ref{eq:H_rho_theta}) of the spring pendulum is given by
\begin{equation} \label{eq:E_distribution}
	E_T = E_S + E_P + E_C.
\end{equation}

The coupling energy term defined by (\ref{eq:EC}) is very suitable for the limit cases the system may present. Supposing that only the spring-mass system moves in time, whereas the pendular motion is suppressed, we have $\theta=0$, $p_{\theta}=0$, and $E_C=f$ (constant). On the other hand, if the spring-mass holds still under the action of gravity in the vertical position, it can be viewed as a rod of fixed length $l$. In this case, only the pendulum moves in time with $\rho=f$, $p_{\rho}=0$, and $E_C=f$ (constant).

For these limit cases, where only the spring-mass or the pendulum moves, the coupling energy term $E_C$ remains constant and equals $f$. We point out that the value of this constant depends on the referential chosen for the potential energy due to gravity $V_g$. In our definitions, we chose $V_g=0$ for $y=0$, and thus we have $E_C=f$. If one chooses $V_g=0$ for $y=-l$, which corresponds to the stable equilibrium position of the system, then $E_C=0$ for the limit cases described above. 

Although $E_C=0$ for the limit cases, when $V_g=0$ for $y=-l$, the position of the referential $V_g=0$ varies with $l$, and consequently with the parameter $f$. For this reason, the referential $V_g=0$ for $y=-l$ is not suitable for the analysis we carry out. Throughout this paper, we work with $V_g=0$ for $y=0$, which is a fixed referential that does not vary with any parameter.

In the next section, we numerically integrate the nonlinear equations of motion for values of energy and control parameter that make the system not tractable analytically. We analyze the dynamics of the system as a whole, including all kinds of trajectories it may present: periodic, quasi-periodic and chaotic orbits.

\section{Results and discussion}
\label{sec:ResultsDiscussion}

\subsection{Single trajectories}
\label{sec:SingleTrajectories}

\begin{figure*}[!tb]
	\centering
	\includegraphics{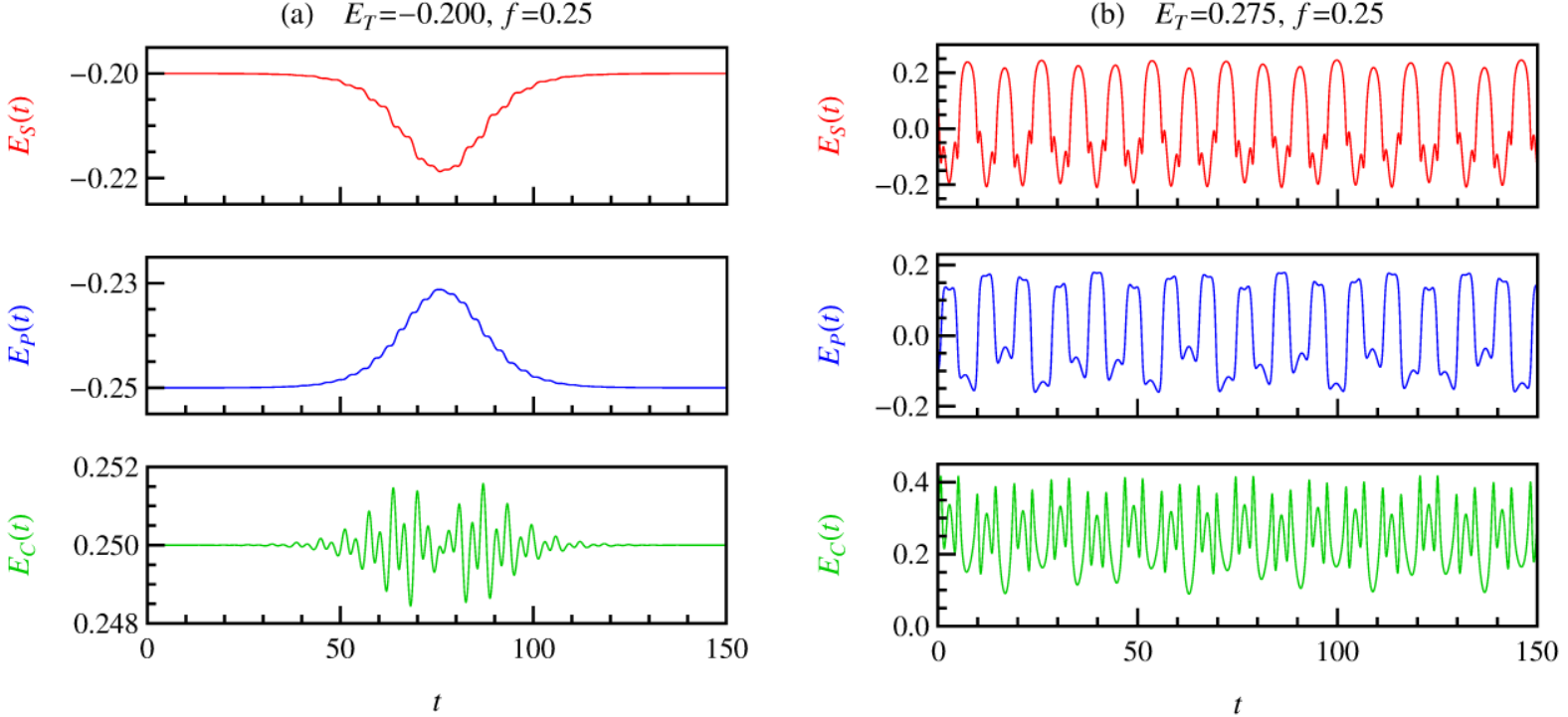}
	\caption{Time evolution of the spring, pendulum and coupling energy terms for the trajectories depicted in Figure \ref{fig:TimeSeries}.}
	\label{fig:EnergyTimeSeries}
\end{figure*}

Most of the publications about the spring pendulum consider the behavior of quasi-periodic trajectories and describe the system using analytical approximations for small angles and low total energy that lead to the parametric resonance condition \cite{Vitt1933,Kane_JAM1968,Tselman_JAMM1970,Rusbridge_AmJP1980,Breitenberger_JMP1981,Lai_AmJP1984}. In this paper, we obtained exact analytical expressions that describe the energy distribution for all kinds of trajectory the system may present: periodic, quasi-periodic and chaotic trajectories.

The analytical expressions (\ref{eq:ES})-(\ref{eq:EC}) for the energy terms are not restricted to specific configurations of the system. They are valid for all values of total energy $E_T$, parameter $f$, weak and strong coupling, small and large oscillations for the $(\rho, \theta)$ coordinates. Using these expressions, we analyze the energy distribution for any individual trajectory of the spring pendulum.

Figure \ref{fig:EnergyTimeSeries} depicts the time evolution of the energy terms (\ref{eq:ES})-(\ref{eq:EC}) for the individual trajectories represented in Figure \ref{fig:TimeSeries}. For both panels of Figure \ref{fig:EnergyTimeSeries}, the total energy of the system remains constant, whereas the energy terms $E_S$, $E_P$ and $E_C$ vary in time.

For $E_T = -0.200$ and $f = 0.25$ as in Figure \ref{fig:EnergyTimeSeries}.(a), the system is close to the parametric resonance. This is the limit case we have described in which either the spring or the pendulum moves at a time. When only the spring-mass or the pendulum moves, all the energy terms (\ref{eq:ES})-(\ref{eq:EC}) remain constant, as can be seen in Figure \ref{fig:EnergyTimeSeries}.(a). When the spring-mass and the pendulum motions exchange energy, the coupling energy term $E_C$ oscillates, causing small oscillations and energy transfer between the spring and pendulum energy terms.

Figure \ref{fig:EnergyTimeSeries}.(b) shows the energy terms (\ref{eq:ES})-(\ref{eq:EC}) for a quasi-periodic trajectory. This kind of trajectory is representative for most of the configurations the system may present with different values of $E_T$ and $f$. In Figure \ref{fig:EnergyTimeSeries}.(b), both the spring and the pendulum move constantly. In this case, all the energy terms oscillate regularly as the spring and the pendulum motions exchange energy.

The energy distribution we propose, including a term due to the nonlinear coupling, reveals new aspects of the spring pendulum dynamics. As can be seen from Figure \ref{fig:EnergyTimeSeries}, the analytical expressions (\ref{eq:ES})-(\ref{eq:EC}) allow us to analyze how the energy is transferred between the spring-mass and the pendulum like motions, and how the two kinds of movement are coupled. The energy distribution we propose introduces a new approach to the study of spring pendulums and other systems with nonlinear coupling.

\subsection{Phase space statistics}
\label{sec:PhaseSpaceStatistics}

The trajectories of the spring pendulum are restricted to a three-dimensional surface delimited by the constant total energy (\ref{eq:Hxy}) in the four-dimensional phase space $(x, y, p_x, p_y)$. To better visualize the trajectories, we use a Poincar\'e section, i.e. we consider the intersections of the trajectories with the plane $q_2 = (y+l)/l = 0$ in dimensionless coordinates, and we plot the points whenever its associated momentum is positive ($p_2 = dq_2/dt > 0$). In the Poincar\'e section, we represent the dimensionless coordinate $q_1 = x/l$ and its associated momentum $p_1 = dq_1/dt$. This Poincar\'e section contains the stable equilibrium position $(x,y)=(0,-l)$, and it exhibits the different kinds of behavior the system may present for a fixed value of total energy and parameter $f$.
\begin{figure*}[!tb]
    \centering
    \includegraphics{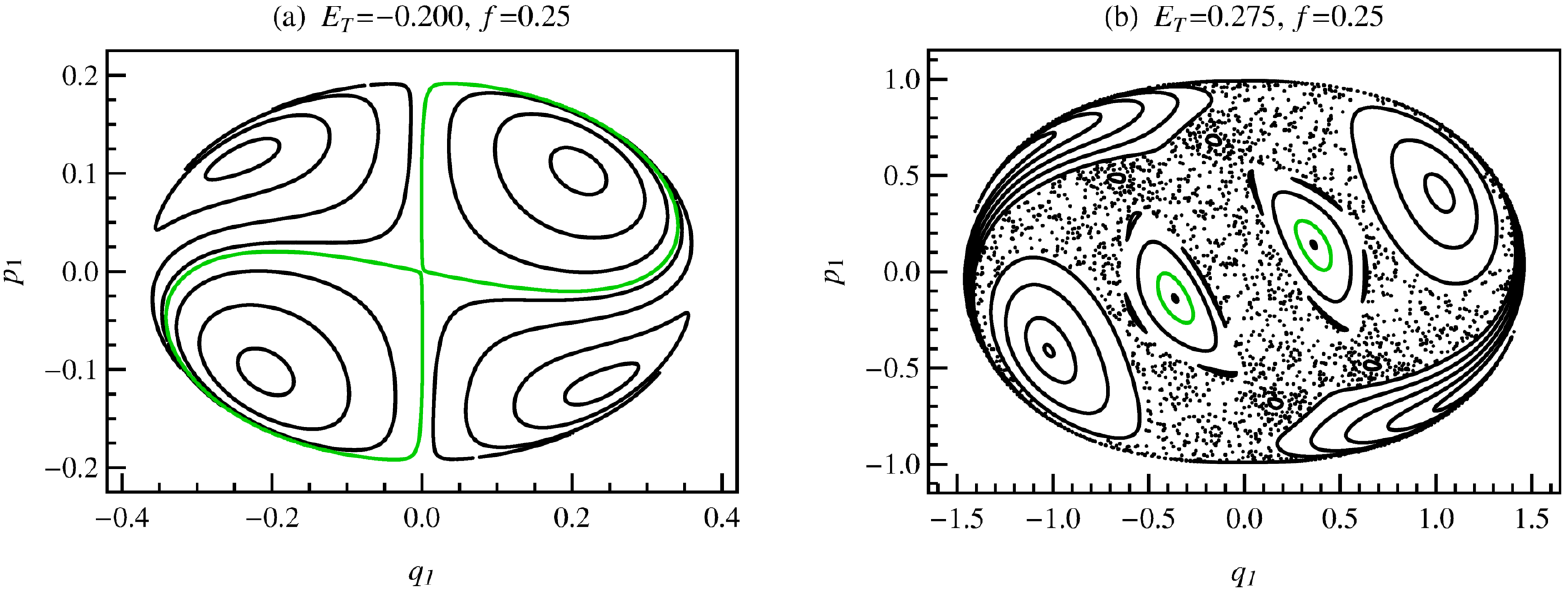}
    \caption{(Color online) Poincar\'e sections showing the different kinds of trajectory presented by the system for $f=0.25$ and (a) $E_T=-0.200$, (b) $E_T=0.275$. The trajectories in green (light grey) correspond to those depicted in Figure \ref{fig:TimeSeries}.}
    \label{fig:PoincareSections}
\end{figure*}

Figure \ref{fig:PoincareSections} shows the Poincar\'e sections of the spring pendulum for the same parameters used in Figure \ref{fig:TimeSeries}. The green (light grey) trajectories in Figure \ref{fig:PoincareSections} correspond to those depicted in Figure \ref{fig:TimeSeries}. For the same value of total energy and $f$, the trajectories in black show us that the system may present regular or chaotic behavior according to the initial conditions. For $E_T=-0.200$ and $f=0.25$ as in Figure \ref{fig:PoincareSections}.(a), the Poincar\'e section of the system is regular. Increasing the value of $E_T$, the Poincar\'e section presents regular islands immersed in a chaotic sea as can be seen in Figure \ref{fig:PoincareSections}.(b). Moreover, in Figure \ref{fig:PoincareSections}.(b), the period-two islands (inner islands) represent pendulum oscillations around the equilibrium position, whereas the period-one islands (outer islands) represent full pendulum rotation around the pivot.

The energy distribution is different for invariant tori, resonant islands and chaotic trajectories. The kind of trajectory that predominates in phase space depends on the total energy and the parameter $f$, as can be seen in Figure \ref{fig:PoincareSections}. Therefore, to understand how the energy distribution varies with $E_T$ and $f$, we consider a great number of trajectories to reproduce all the possible behaviors in phase space and all the dynamical properties the system presents. We calculate the average energy terms for the system and we show that they vary regularly with $E_T$ and $f$.

For each value of $E_T$ and $f$, we choose around 20000 initial conditions evenly distributed in an elliptical grid that covers the entire Poincar\'e section $p_1 \times q_1$, with $q_2 = 0$ and $p_2 > 0$. These initial conditions correspond to all kinds of trajectories the spring pendulum may present: regular and chaotic orbits, pendulum oscillation around the equilibrium position, pendulum rotation around the pivot, arbitrary amplitude oscillations for the spring. For each initial condition, we integrate the Hamilton's equations to obtain the time evolution of the trajectory, and we evaluate the temporal average energy terms in the time interval $t=[0,500]$. Considering all the initial conditions, we calculate the average energy terms $\overline{E}_i$ for the whole phase space, obtaining results that are both spatial and temporal averages. We then normalize the average energies for each component (spring, pendulum and coupling) to the interval $[0,1]$.

It is important to notice that 20000 initial conditions are sufficient to accurately describe the features of phase space and the average energy terms we compute. If we work with a different set of 20000 initial conditions, the maximum difference in the normalized average energy terms is on the order of $10^{-3}$. Doubling the number of initial conditions also produces a maximum difference on the order of $10^{-3}$ in the results.
\begin{figure*}[!tb]
	\centering
	\includegraphics{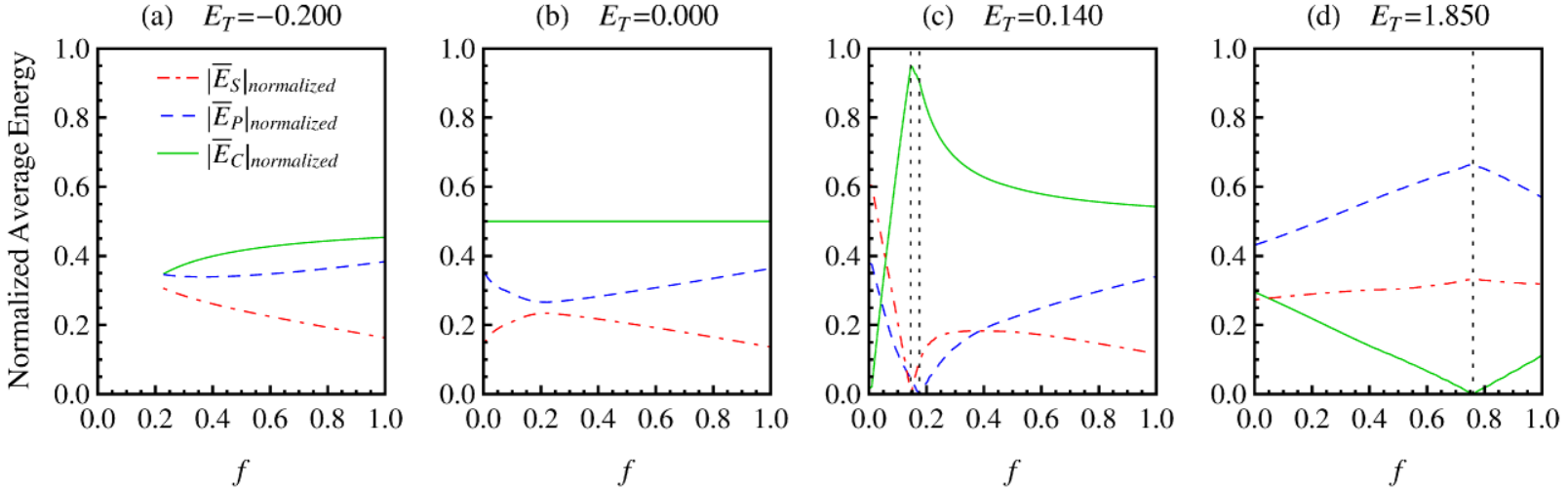}
	\caption{(Color online) Normalized average energy as a function of the parameter $f$ for different values of the total energy $E_T$. The red (dot dashed) curves correspond to the normalized average energy of the spring term, the blue (dashed) curves represent the pendulum energy term, and the green (solid) curves represent the coupling energy term. The vertical dotted lines in black indicate the position of minimum and maximum values.}
	\label{fig:Emedia}
\end{figure*}

Figure \ref{fig:Emedia} shows the normalized average energy terms $|\overline{E}_S|_{N}$, $|\overline{E}_P|_{N}$ and $|\overline{E}_C|_{N}$ for different values of the total energy $E_T$. From this figure, we observe that the average energy terms follow a regular pattern as we increase the value of $E_T$. The behavior of the average energy terms according to $E_T$ can also be observed in the video included in the Supplementary Material of this paper. In the video, we show the energy terms $|\overline{E}_i|_{N}$ for $E_T = ]-0.50, 3.00]$, with $E_T = -0.50$ the minimum energy the system may present according to condition $E_T \geqslant f^2/2 - f$, i.e. the total energy must be greater than the energy of the system in its stable equilibrium position ($\rho=l$, $\theta=0$, $p_{\rho}=0$, and $p_{\theta}=0$).

The spring pendulum presents an order-chaos-order transition as we increase the total energy or the parameter $f$ from their minimum values \cite{Yepez_PLA1990,Cuerno_AmJP92,Gonzalez_EJP94,Weele_PhysA1996}. For low values of $E_T$ or $f$, the system is regular. Increasing the total energy or the parameter $f$, the system starts to present chaotic trajectories. The area covered by chaos expands with $E_T$ and $f$, but at some point it begins to diminish. For sufficiently large values of $E_T$ or $f$, the system becomes regular once again. For $E_T > 3.00$, the system is regular irrespective of the value of $f$, and its behavior does not change much with increasing energies. Therefore, in the interval $E_T = ]-0.50, 3.00]$, we comprise all the dynamical features the system may present: regular and chaotic behavior, resonant trajectories, pendulum oscillation around the equilibrium position, and pendulum rotation.

For negative values of total energy as in Figure \ref{fig:Emedia}.(a), the average coupling energy term is always greater than the average spring and pendulum energy terms, whereas the average spring energy term is the smallest one. From Figure \ref{fig:Emedia}.(a), we also notice that for $E_T<0$, some values of the parameter $f$ are not allowed because they do not comply with the condition of minimum energy.

When the total energy of the system is null, the average coupling energy term presents a very interesting behavior as can be seen in Figure \ref{fig:Emedia}.(b). For $E_T=0$, $|\overline{E}_C|_{N}$ is constant and equal to $0.5$ for all values of $f$. It means that, when $E_T=0$, the coupling energy term concentrates, on average, exactly half of the system energy and it does not depend on the parameter $f$.

For low positive values of total energy as in Figure \ref{fig:Emedia}.(c), the average coupling energy term is generally higher than the spring and pendulum energy terms. For most values of $f$, we have $|\overline{E}_C|_{N} > 0.5$, indicating a strong coupling in the system. When the coupling is strong, the spring and pendulum like motions exchange a great amount of energy and it is difficult to distinguish the two types of movement.

For intermediate positive values of the total energy as in Figure \ref{fig:Emedia}.(d), the average coupling energy term is lower than the average spring and pendulum energy terms, whereas the average pendulum energy term is the highest one. For such values of $E_T$, $|\overline{E}_C|_{N} < 0.5$, and $|\overline{E}_C|_{N}$ is null for specific values of $f$, which indicates a weak coupling in the system. The position of $|\overline{E}_C|_{N} = 0$ varies regularly with $E_T$ and $f$. Using the nonlinear least-squares (NLLS) Marquardt-Levenberg algorithm, we fit the numerical data and obtain a second order polynomial $f = 0.19E_T^2 + 0.054E_T - 0.0034$ that describes the position of $|\overline{E}_C|_{N} = 0$ as a function of $E_T$ and $f$. When $|\overline{E}_C|_{N} = 0$, the spring and pendulum energy terms are maximum, and they concentrate, on average, all the energy of the system.
\begin{figure}[!tb]
	\centering
	\includegraphics{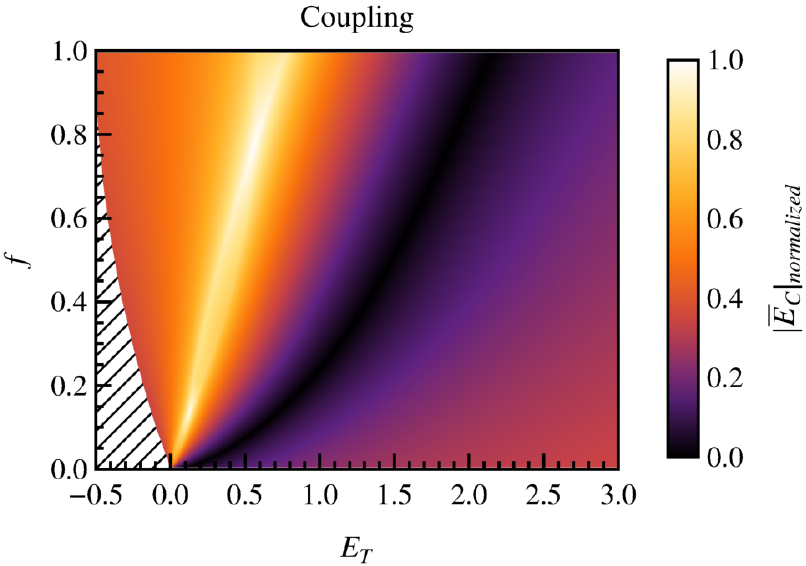}
	\caption{(Color online) Normalized average coupling energy term as a function of $E_T$ and $f$. The hatched area is not allowed because these values of $E_T$ and $f$ do not comply with the condition of minimum energy $E_T \geqslant f^2/2 - f$.}
	\label{fig:EmediaCoup}
\end{figure}

Figure \ref{fig:EmediaCoup} shows the normalized average coupling energy term as a function of the total energy $E_T$ and the parameter $f$. For negative and low positive values of the total energy, the coupling in the system is stronger, as indicated by the orange (light grey) color in Figure \ref{fig:EmediaCoup}. The average coupling energy term reaches its maximum values ($|\overline{E}_C|_{N} \simeq 1$) in the white region of the picture. In this situation, all the energy of the system, on average, is concentrated in the coupling energy term, whereas the average spring and pendulum energy terms vanish.

For intermediate positive values of the total energy, the coupling in the system becomes weak and reaches its minimum values ($|\overline{E}_C|_{N} \simeq 0$) in the black region of Figure \ref{fig:EmediaCoup}. This black region is centered on the second order polynomial $f = 0.19E_T^2 + 0.054E_T - 0.0034$ we mentioned. After the black region in Figure \ref{fig:EmediaCoup} for which $|\overline{E}_C|_{N} \simeq 0$, the average coupling energy term starts to increase again. The purple (intermediate grey) color in the picture indicates that the coupling in the system is moderate when the total energy $E_T$ is high. For these values of $E_T$, the pendular motion dominates the system, whereas the average spring energy term is the smallest one.

When the coupling in the system is strong ($|\overline{E}_C|_{N} \simeq 1$), the average spring and pendulum energy terms reach their minimum values. On the other hand, for weak coupling ($|\overline{E}_C|_{N} \simeq 0$), the average spring and pendulum energy terms are maximum. In the Supplementary Material of this paper, we show two pictures similar to Figure \ref{fig:EmediaCoup} representing the normalized average spring and pendulum energy terms as a function of $E_T$ and $f$. In these pictures, it is possible to identify the regions where the average spring and pendulum energy terms dominate the dynamics and where these terms reach their minimum values.

The procedure we propose is valid for any configuration of the system. Thus, it allows us to analyze a great number of trajectories and to investigate the average energy distribution in phase space according to the total energy $E_T$ and the parameter $f$ that accounts for the physical characteristics of the system. With this procedure, we present a new way to study the nonlinear coupling and the internal energy distribution among the system components.

\section{Conclusions}
\label{sec:Conclusions}

We investigated nonlinear coupled systems by considering the total energy distributed among the system components and their coupling. As an example, we analyzed a spring pendulum, which represents a paradigm for these systems. In the spring pendulum, the spring and pendulum like motions are coupled. Therefore, the system presents a coupling energy term that mediates the energy exchanges between the two kinds of movement.

We considered the total energy of the spring pendulum distributed among three terms: spring, pendulum and coupling. We obtained analytical expressions for the three energy terms. These expressions are valid for any value of total energy, system parameters, weak and strong coupling, small and large amplitude oscillations for the spring and pendular movements. We verified that our analytical expressions accurately describe the energy exchanges that occur between the spring and pendulum like movements, including the cases of parametric resonance, regular and chaotic orbits.

We used our definition for the energy terms to study the global behavior of the spring pendulum. To do so, we evaluated the average energy terms for a great number of trajectories throughout the phase space, obtaining results that are both temporal and spatial averaged. We verified that the average energy distribution varies regularly as a function of the total energy and a parameter that accounts for the physical characteristics of the system.

From the average coupling energy term, we identified regions of strong and weak coupling in the parameter space of the system. When the coupling is strong, the spring and pendulum exchange a great amount of energy. The two subsystems behave as a unique new system and, most of the time, it is difficult to identify the individual spring-mass and pendular like motions.

For some regions in the parameter space, the coupling in the system is weak. The spring and the pendulum slightly interfere in each other motion and it is easy to identify the two different kinds of movement in certain periods of time. We also observed regions of moderate coupling in the parameter space. In this case, the pendulum energy term dominates the dynamics of the system, whereas the spring energy term is the lowest one.

The new approach we proposed in this paper, considering a coupling energy term, allowed us to observe new features of the spring pendulum dynamics, and to determine how the coupling mediates internal energy exchanges between the different kinds of movement the system may present. By distributing the total energy of the spring pendulum among its two subsystems and their coupling, we determined which subsystem dominates, on average, the dynamics of the system. We also verified that the energy distribution, and the dominant subsystem, varies according to the total energy $E_T$ and the parameter $f$ that accounts for the system physical characteristics. For some values of $E_T$ and $f$, either the spring-mass or the pendulum subsystem dominates the dynamics. For other parameters values, the dynamics is dominated by the coupling, meaning that the two subsystems exchange energy constantly, and it is difficult to distinguish the individual spring and pendulum movements.

This kind of analysis is useful to the large number of mechanical devices that use the spring pendulum as a component \cite{Anh2007}, as well as to the nonlinear coupled systems that use the spring pendulum as a model to describe their dynamics. Among these, we may cite different mechanical systems \cite{Holmes2006,Wang2011,Castillo-Rivera2017}, the orbits of celestial bodies \cite{Contopoulos_AJ1963,Hori_ASJ1966,Broucke_CM1973,Hitzl_CM1975}, the classical analogue for the vibrational modes of triatomic molecules producing the Fermi resonance in the infrared and Raman spectra \cite{Ramaneffekt1931,Amat_JMS1965,Jacob_JPB1978}, and the nonlinear interaction between light waves \cite{Armstrong_PR1962}.

We point out that the methods we developed are not restricted to the analysis of spring pendulums. They may be applied to other nonlinear coupled systems for which it is possible to identify the energy terms associated with each subsystem. Following the strategy we presented, one obtains analytical expressions for the energy terms associated with each subsystem and their coupling, verifying the configurations that lead to weak and strong coupling in the system, how the coupling mediates internal energy exchanges, and whether the coupled system dynamics is dominated by one of the subsystems or by their coupling.

One example of possible application for the proposed procedure is the study of wave coupling in plasma physics \cite{Sagdeev1969,Ritz1986,Ritz1988,Horton2012}. The methods we presented can describe how the two waves are coupled, how the energy is transferred from one wave to the other, and which wave concentrates more energy, on average, according to the system parameters. This approach provides new perspectives and contributes to a better understanding about the coupled system dynamics, the coupling among its components, and how the energy distribution regulates the behavior of the nonlinear system.

\section*{Supplementary Material}

See Supplementary Material for the parameter spaces representing the normalized average spring and pendulum energy terms as a function of the total energy $E_T$ and the parameter $f$. We also present a video that shows how the normalized average energy terms vary according to these parameters.

\begin{acknowledgments}
We thank Dr. Kai Ullmann and Prof. Dr. Alfredo M. Ozorio de Almeida for the discussions that contributed to the present work. Funding: This work was supported by the Brazilian scientific agencies: S\~ao Paulo Research Foundation (FAPESP) [grant numbers 2015/05186-0, 2011/19296-1], Conselho Nacional de Desenvolvimento Cient\'ifico e Tecnol\'ogico (CNPq) [grant numbers 457030/2014-3, 157317/2015-3], and Coordena\c{c}\~ao de Aperfei\c{c}oamento de Pessoal de N\'ivel Superior (Capes).
\end{acknowledgments}

\bibliographystyle{aipnum4-1}
\bibliography{ElastPend_bib}

\begin{thebibliography}{56}%
\makeatletter
\providecommand \@ifxundefined [1]{%
 \@ifx{#1\undefined}
}%
\providecommand \@ifnum [1]{%
 \ifnum #1\expandafter \@firstoftwo
 \else \expandafter \@secondoftwo
 \fi
}%
\providecommand \@ifx [1]{%
 \ifx #1\expandafter \@firstoftwo
 \else \expandafter \@secondoftwo
 \fi
}%
\providecommand \natexlab [1]{#1}%
\providecommand \enquote  [1]{``#1''}%
\providecommand \bibnamefont  [1]{#1}%
\providecommand \bibfnamefont [1]{#1}%
\providecommand \citenamefont [1]{#1}%
\providecommand \href@noop [0]{\@secondoftwo}%
\providecommand \href [0]{\begingroup \@sanitize@url \@href}%
\providecommand \@href[1]{\@@startlink{#1}\@@href}%
\providecommand \@@href[1]{\endgroup#1\@@endlink}%
\providecommand \@sanitize@url [0]{\catcode `\\12\catcode `\$12\catcode
  `\&12\catcode `\#12\catcode `\^12\catcode `\_12\catcode `\%12\relax}%
\providecommand \@@startlink[1]{}%
\providecommand \@@endlink[0]{}%
\providecommand \url  [0]{\begingroup\@sanitize@url \@url }%
\providecommand \@url [1]{\endgroup\@href {#1}{\urlprefix }}%
\providecommand \urlprefix  [0]{URL }%
\providecommand \Eprint [0]{\href }%
\providecommand \doibase [0]{http://dx.doi.org/}%
\providecommand \selectlanguage [0]{\@gobble}%
\providecommand \bibinfo  [0]{\@secondoftwo}%
\providecommand \bibfield  [0]{\@secondoftwo}%
\providecommand \translation [1]{[#1]}%
\providecommand \BibitemOpen [0]{}%
\providecommand \bibitemStop [0]{}%
\providecommand \bibitemNoStop [0]{.\EOS\space}%
\providecommand \EOS [0]{\spacefactor3000\relax}%
\providecommand \BibitemShut  [1]{\csname bibitem#1\endcsname}%
\let\auto@bib@innerbib\@empty
\bibitem [{\citenamefont {Sagdeev}\ and\ \citenamefont
  {Galeev}(1969)}]{Sagdeev1969}%
  \BibitemOpen
  \bibfield  {author} {\bibinfo {author} {\bibfnamefont {R.~Z.}\ \bibnamefont
  {Sagdeev}}\ and\ \bibinfo {author} {\bibfnamefont {A.~A.}\ \bibnamefont
  {Galeev}},\ }\href@noop {} {\emph {\bibinfo {title} {{Nonlinear Plasma
  Theory}}}},\ edited by\ \bibinfo {editor} {\bibfnamefont {T.~M.}\
  \bibnamefont {O'Neil}}\ and\ \bibinfo {editor} {\bibfnamefont {D.~L.}\
  \bibnamefont {Book}}\ (\bibinfo  {publisher} {Benjamin},\ \bibinfo {address}
  {New York},\ \bibinfo {year} {1969})\ \bibinfo {note} {chap. I}\BibitemShut
  {NoStop}%
\bibitem [{\citenamefont {Ritz}\ and\ \citenamefont {Powers}(1986)}]{Ritz1986}%
  \BibitemOpen
  \bibfield  {author} {\bibinfo {author} {\bibfnamefont {C.~P.}\ \bibnamefont
  {Ritz}}\ and\ \bibinfo {author} {\bibfnamefont {E.~J.}\ \bibnamefont
  {Powers}},\ }\href@noop {} {\bibfield  {journal} {\bibinfo  {journal}
  {Physica D}\ }\textbf {\bibinfo {volume} {20}},\ \bibinfo {pages} {320}
  (\bibinfo {year} {1986})}\BibitemShut {NoStop}%
\bibitem [{\citenamefont {Ritz}\ \emph {et~al.}(1988)\citenamefont {Ritz},
  \citenamefont {Powers}, \citenamefont {Rhodes}, \citenamefont {Bengtson},
  \citenamefont {Gentle}, \citenamefont {Lin}, \citenamefont {Phillips},
  \citenamefont {Wootton}, \citenamefont {Brower}, \citenamefont {{Luhmann
  Jr.}}, \citenamefont {Peebles}, \citenamefont {Schoch},\ and\ \citenamefont
  {Hickok}}]{Ritz1988}%
  \BibitemOpen
  \bibfield  {author} {\bibinfo {author} {\bibfnamefont {C.~P.}\ \bibnamefont
  {Ritz}}, \bibinfo {author} {\bibfnamefont {E.~J.}\ \bibnamefont {Powers}},
  \bibinfo {author} {\bibfnamefont {T.~L.}\ \bibnamefont {Rhodes}}, \bibinfo
  {author} {\bibfnamefont {R.~D.}\ \bibnamefont {Bengtson}}, \bibinfo {author}
  {\bibfnamefont {K.~W.}\ \bibnamefont {Gentle}}, \bibinfo {author}
  {\bibfnamefont {H.}~\bibnamefont {Lin}}, \bibinfo {author} {\bibfnamefont
  {P.~E.}\ \bibnamefont {Phillips}}, \bibinfo {author} {\bibfnamefont {A.~J.}\
  \bibnamefont {Wootton}}, \bibinfo {author} {\bibfnamefont {D.~L.}\
  \bibnamefont {Brower}}, \bibinfo {author} {\bibfnamefont {N.~C.}\
  \bibnamefont {{Luhmann Jr.}}}, \bibinfo {author} {\bibfnamefont {W.~A.}\
  \bibnamefont {Peebles}}, \bibinfo {author} {\bibfnamefont {P.~M.}\
  \bibnamefont {Schoch}}, \ and\ \bibinfo {author} {\bibfnamefont {R.~L.}\
  \bibnamefont {Hickok}},\ }\href@noop {} {\bibfield  {journal} {\bibinfo
  {journal} {Review of Scientific Instruments}\ }\textbf {\bibinfo {volume}
  {59}},\ \bibinfo {pages} {1739} (\bibinfo {year} {1988})}\BibitemShut
  {NoStop}%
\bibitem [{\citenamefont {Horton}(2012)}]{Horton2012}%
  \BibitemOpen
  \bibfield  {author} {\bibinfo {author} {\bibfnamefont {W.}~\bibnamefont
  {Horton}},\ }\href@noop {} {\emph {\bibinfo {title} {Turbulent transport in
  magnetized plasmas}}}\ (\bibinfo  {publisher} {World Scientific},\ \bibinfo
  {address} {Singapore},\ \bibinfo {year} {2012})\BibitemShut {NoStop}%
\bibitem [{\citenamefont {Wiesenfeld}\ \emph {et~al.}(1990)\citenamefont
  {Wiesenfeld}, \citenamefont {Bracikowski}, \citenamefont {James},\ and\
  \citenamefont {Roy}}]{Wiesenfeld1990}%
  \BibitemOpen
  \bibfield  {author} {\bibinfo {author} {\bibfnamefont {K.}~\bibnamefont
  {Wiesenfeld}}, \bibinfo {author} {\bibfnamefont {C.}~\bibnamefont
  {Bracikowski}}, \bibinfo {author} {\bibfnamefont {G.}~\bibnamefont {James}},
  \ and\ \bibinfo {author} {\bibfnamefont {R.}~\bibnamefont {Roy}},\
  }\href@noop {} {\bibfield  {journal} {\bibinfo  {journal} {Physical Review
  Letters}\ }\textbf {\bibinfo {volume} {65}},\ \bibinfo {pages} {1749}
  (\bibinfo {year} {1990})}\BibitemShut {NoStop}%
\bibitem [{\citenamefont {Kozyreff}, \citenamefont {Vladimirov},\ and\
  \citenamefont {Mandel}(2000)}]{Kozyreff2000}%
  \BibitemOpen
  \bibfield  {author} {\bibinfo {author} {\bibfnamefont {G.}~\bibnamefont
  {Kozyreff}}, \bibinfo {author} {\bibfnamefont {A.~G.}\ \bibnamefont
  {Vladimirov}}, \ and\ \bibinfo {author} {\bibfnamefont {P.}~\bibnamefont
  {Mandel}},\ }\href@noop {} {\bibfield  {journal} {\bibinfo  {journal}
  {Physical Review Letters}\ }\textbf {\bibinfo {volume} {85}},\ \bibinfo
  {pages} {3809} (\bibinfo {year} {2000})}\BibitemShut {NoStop}%
\bibitem [{\citenamefont {Zamora-Munt}\ \emph {et~al.}(2010)\citenamefont
  {Zamora-Munt}, \citenamefont {Masoller}, \citenamefont {Garcia-Ojalvo},\ and\
  \citenamefont {Roy}}]{Zamora-Munt2010}%
  \BibitemOpen
  \bibfield  {author} {\bibinfo {author} {\bibfnamefont {J.}~\bibnamefont
  {Zamora-Munt}}, \bibinfo {author} {\bibfnamefont {C.}~\bibnamefont
  {Masoller}}, \bibinfo {author} {\bibfnamefont {J.}~\bibnamefont
  {Garcia-Ojalvo}}, \ and\ \bibinfo {author} {\bibfnamefont {R.}~\bibnamefont
  {Roy}},\ }\href@noop {} {\bibfield  {journal} {\bibinfo  {journal} {Physical
  Review Letters}\ }\textbf {\bibinfo {volume} {105}},\ \bibinfo {pages}
  {264101} (\bibinfo {year} {2010})}\BibitemShut {NoStop}%
\bibitem [{\citenamefont {Winfree}(1980)}]{Winfree1980}%
  \BibitemOpen
  \bibfield  {author} {\bibinfo {author} {\bibfnamefont {A.}~\bibnamefont
  {Winfree}},\ }\href@noop {} {\emph {\bibinfo {title} {The Geometry of
  Biological Time}}}\ (\bibinfo  {publisher} {Springer},\ \bibinfo {address}
  {New York},\ \bibinfo {year} {1980})\BibitemShut {NoStop}%
\bibitem [{\citenamefont {Kuramoto}(1984)}]{Kuramoto1984}%
  \BibitemOpen
  \bibfield  {author} {\bibinfo {author} {\bibfnamefont {Y.}~\bibnamefont
  {Kuramoto}},\ }\href@noop {} {\emph {\bibinfo {title} {Chemical Oscillations,
  Waves and Turbulence}}}\ (\bibinfo  {publisher} {Springer},\ \bibinfo
  {address} {Berlin},\ \bibinfo {year} {1984})\BibitemShut {NoStop}%
\bibitem [{\citenamefont {Strogatz}\ and\ \citenamefont
  {Stewart}(1993)}]{Strogatz1993}%
  \BibitemOpen
  \bibfield  {author} {\bibinfo {author} {\bibfnamefont {S.~H.}\ \bibnamefont
  {Strogatz}}\ and\ \bibinfo {author} {\bibfnamefont {I.}~\bibnamefont
  {Stewart}},\ }\href@noop {} {\bibfield  {journal} {\bibinfo  {journal}
  {Scientific American}\ }\textbf {\bibinfo {volume} {269}},\ \bibinfo {pages}
  {102} (\bibinfo {year} {1993})}\BibitemShut {NoStop}%
\bibitem [{\citenamefont {Bressloff}, \citenamefont {Coombes},\ and\
  \citenamefont {de~Souza}(1997)}]{Bressloff1997}%
  \BibitemOpen
  \bibfield  {author} {\bibinfo {author} {\bibfnamefont {P.~C.}\ \bibnamefont
  {Bressloff}}, \bibinfo {author} {\bibfnamefont {S.}~\bibnamefont {Coombes}},
  \ and\ \bibinfo {author} {\bibfnamefont {B.}~\bibnamefont {de~Souza}},\
  }\href@noop {} {\bibfield  {journal} {\bibinfo  {journal} {Physical Review
  Letters}\ }\textbf {\bibinfo {volume} {79}},\ \bibinfo {pages} {2791}
  (\bibinfo {year} {1997})}\BibitemShut {NoStop}%
\bibitem [{\citenamefont {Newman}(2010)}]{Newman2010}%
  \BibitemOpen
  \bibfield  {author} {\bibinfo {author} {\bibfnamefont {M.~E.~J.}\
  \bibnamefont {Newman}},\ }\href@noop {} {\emph {\bibinfo {title} {Networks:
  An Introduction}}}\ (\bibinfo  {publisher} {Oxford University Press},\
  \bibinfo {address} {Oxford},\ \bibinfo {year} {2010})\BibitemShut {NoStop}%
\bibitem [{\citenamefont {Abbott}\ and\ \citenamefont {van
  Vreeswijk}(1993)}]{Abbott1993}%
  \BibitemOpen
  \bibfield  {author} {\bibinfo {author} {\bibfnamefont {L.~F.}\ \bibnamefont
  {Abbott}}\ and\ \bibinfo {author} {\bibfnamefont {C.}~\bibnamefont {van
  Vreeswijk}},\ }\href@noop {} {\bibfield  {journal} {\bibinfo  {journal}
  {Physical Review E}\ }\textbf {\bibinfo {volume} {48}},\ \bibinfo {pages}
  {1483} (\bibinfo {year} {1993})}\BibitemShut {NoStop}%
\bibitem [{\citenamefont {Collins}, \citenamefont {Chow},\ and\ \citenamefont
  {Imhoff}(1995)}]{Collins1995}%
  \BibitemOpen
  \bibfield  {author} {\bibinfo {author} {\bibfnamefont {J.~J.}\ \bibnamefont
  {Collins}}, \bibinfo {author} {\bibfnamefont {C.~C.}\ \bibnamefont {Chow}}, \
  and\ \bibinfo {author} {\bibfnamefont {T.~T.}\ \bibnamefont {Imhoff}},\
  }\href@noop {} {\bibfield  {journal} {\bibinfo  {journal} {Nature}\ }\textbf
  {\bibinfo {volume} {376}},\ \bibinfo {pages} {236} (\bibinfo {year}
  {1995})}\BibitemShut {NoStop}%
\bibitem [{\citenamefont {Joya}, \citenamefont {Atencia},\ and\ \citenamefont
  {Sandoval}(2002)}]{Joya2002}%
  \BibitemOpen
  \bibfield  {author} {\bibinfo {author} {\bibfnamefont {G.}~\bibnamefont
  {Joya}}, \bibinfo {author} {\bibfnamefont {M.~A.}\ \bibnamefont {Atencia}}, \
  and\ \bibinfo {author} {\bibfnamefont {F.}~\bibnamefont {Sandoval}},\
  }\href@noop {} {\bibfield  {journal} {\bibinfo  {journal} {Neurocomputing}\
  }\textbf {\bibinfo {volume} {43}},\ \bibinfo {pages} {219} (\bibinfo {year}
  {2002})}\BibitemShut {NoStop}%
\bibitem [{\citenamefont {Wang}, \citenamefont {Wang},\ and\ \citenamefont
  {Liu}(2010)}]{Wang2010}%
  \BibitemOpen
  \bibfield  {author} {\bibinfo {author} {\bibfnamefont {Z.}~\bibnamefont
  {Wang}}, \bibinfo {author} {\bibfnamefont {Y.}~\bibnamefont {Wang}}, \ and\
  \bibinfo {author} {\bibfnamefont {Y.}~\bibnamefont {Liu}},\ }\href@noop {}
  {\bibfield  {journal} {\bibinfo  {journal} {IEEE Transactions on Neural
  Networks}\ }\textbf {\bibinfo {volume} {21}},\ \bibinfo {pages} {11}
  (\bibinfo {year} {2010})}\BibitemShut {NoStop}%
\bibitem [{\citenamefont {Bolouri}\ and\ \citenamefont
  {Davidson}(2002)}]{Bolouri2002}%
  \BibitemOpen
  \bibfield  {author} {\bibinfo {author} {\bibfnamefont {H.}~\bibnamefont
  {Bolouri}}\ and\ \bibinfo {author} {\bibfnamefont {E.~H.}\ \bibnamefont
  {Davidson}},\ }\href@noop {} {\bibfield  {journal} {\bibinfo  {journal}
  {BioEssays}\ }\textbf {\bibinfo {volume} {24}},\ \bibinfo {pages} {1118}
  (\bibinfo {year} {2002})}\BibitemShut {NoStop}%
\bibitem [{\citenamefont {{De Jong}}(2002)}]{DeJong2002}%
  \BibitemOpen
  \bibfield  {author} {\bibinfo {author} {\bibfnamefont {H.}~\bibnamefont {{De
  Jong}}},\ }\href@noop {} {\bibfield  {journal} {\bibinfo  {journal} {Journal
  of Computational Biology}\ }\textbf {\bibinfo {volume} {9}},\ \bibinfo
  {pages} {67} (\bibinfo {year} {2002})}\BibitemShut {NoStop}%
\bibitem [{\citenamefont {Ren}\ and\ \citenamefont {Cao}(2008)}]{Ren2008}%
  \BibitemOpen
  \bibfield  {author} {\bibinfo {author} {\bibfnamefont {F.}~\bibnamefont
  {Ren}}\ and\ \bibinfo {author} {\bibfnamefont {J.}~\bibnamefont {Cao}},\
  }\href@noop {} {\bibfield  {journal} {\bibinfo  {journal} {Neurocomputing}\
  }\textbf {\bibinfo {volume} {71}},\ \bibinfo {pages} {834} (\bibinfo {year}
  {2008})}\BibitemShut {NoStop}%
\bibitem [{\citenamefont {Ford}(1961)}]{Ford1961}%
  \BibitemOpen
  \bibfield  {author} {\bibinfo {author} {\bibfnamefont {J.}~\bibnamefont
  {Ford}},\ }\href@noop {} {\bibfield  {journal} {\bibinfo  {journal} {Journal
  of Mathematical Physics}\ }\textbf {\bibinfo {volume} {2}},\ \bibinfo {pages}
  {387} (\bibinfo {year} {1961})}\BibitemShut {NoStop}%
\bibitem [{\citenamefont {Jackson}(1963)}]{Jackson1963}%
  \BibitemOpen
  \bibfield  {author} {\bibinfo {author} {\bibfnamefont {E.~A.}\ \bibnamefont
  {Jackson}},\ }\href@noop {} {\bibfield  {journal} {\bibinfo  {journal}
  {Journal of Mathematical Physics}\ }\textbf {\bibinfo {volume} {4}},\
  \bibinfo {pages} {686} (\bibinfo {year} {1963})}\BibitemShut {NoStop}%
\bibitem [{\citenamefont {Gendelman}(2001)}]{Gendelman2001}%
  \BibitemOpen
  \bibfield  {author} {\bibinfo {author} {\bibfnamefont {O.~V.}\ \bibnamefont
  {Gendelman}},\ }\href@noop {} {\bibfield  {journal} {\bibinfo  {journal}
  {Nonlinear Dynamics}\ }\textbf {\bibinfo {volume} {25}},\ \bibinfo {pages}
  {237} (\bibinfo {year} {2001})}\BibitemShut {NoStop}%
\bibitem [{\citenamefont {Vakakis}\ and\ \citenamefont
  {Rand}(2004)}]{Vakakis2004}%
  \BibitemOpen
  \bibfield  {author} {\bibinfo {author} {\bibfnamefont {A.~F.}\ \bibnamefont
  {Vakakis}}\ and\ \bibinfo {author} {\bibfnamefont {R.~H.}\ \bibnamefont
  {Rand}},\ }\href@noop {} {\bibfield  {journal} {\bibinfo  {journal}
  {International Journal of Non-Linear Mechanics}\ }\textbf {\bibinfo {volume}
  {39}},\ \bibinfo {pages} {1079} (\bibinfo {year} {2004})}\BibitemShut
  {NoStop}%
\bibitem [{\citenamefont {Quinn}\ \emph {et~al.}(2008)\citenamefont {Quinn},
  \citenamefont {Gendelman}, \citenamefont {Kerschen}, \citenamefont {Sapsis},
  \citenamefont {Bergman},\ and\ \citenamefont {Vakakis}}]{Quinn2008}%
  \BibitemOpen
  \bibfield  {author} {\bibinfo {author} {\bibfnamefont {D.~D.}\ \bibnamefont
  {Quinn}}, \bibinfo {author} {\bibfnamefont {O.}~\bibnamefont {Gendelman}},
  \bibinfo {author} {\bibfnamefont {G.}~\bibnamefont {Kerschen}}, \bibinfo
  {author} {\bibfnamefont {T.~P.}\ \bibnamefont {Sapsis}}, \bibinfo {author}
  {\bibfnamefont {L.~A.}\ \bibnamefont {Bergman}}, \ and\ \bibinfo {author}
  {\bibfnamefont {A.~F.}\ \bibnamefont {Vakakis}},\ }\href@noop {} {\bibfield
  {journal} {\bibinfo  {journal} {Journal of Sound and Vibration}\ }\textbf
  {\bibinfo {volume} {311}},\ \bibinfo {pages} {1228} (\bibinfo {year}
  {2008})}\BibitemShut {NoStop}%
\bibitem [{\citenamefont {Kovaleva}, \citenamefont {Manevitch},\ and\
  \citenamefont {Manevitch}(2010)}]{Kovaleva2010}%
  \BibitemOpen
  \bibfield  {author} {\bibinfo {author} {\bibfnamefont {A.}~\bibnamefont
  {Kovaleva}}, \bibinfo {author} {\bibfnamefont {L.}~\bibnamefont {Manevitch}},
  \ and\ \bibinfo {author} {\bibfnamefont {E.}~\bibnamefont {Manevitch}},\
  }\href@noop {} {\bibfield  {journal} {\bibinfo  {journal} {Physical Review
  E}\ }\textbf {\bibinfo {volume} {81}},\ \bibinfo {pages} {056215} (\bibinfo
  {year} {2010})}\BibitemShut {NoStop}%
\bibitem [{\citenamefont {Sigalov}\ \emph {et~al.}(2012)\citenamefont
  {Sigalov}, \citenamefont {Gendelman}, \citenamefont {AL-Shudeifat},
  \citenamefont {Manevitch}, \citenamefont {Vakakis},\ and\ \citenamefont
  {Bergman}}]{Sigalov2012}%
  \BibitemOpen
  \bibfield  {author} {\bibinfo {author} {\bibfnamefont {G.}~\bibnamefont
  {Sigalov}}, \bibinfo {author} {\bibfnamefont {O.~V.}\ \bibnamefont
  {Gendelman}}, \bibinfo {author} {\bibfnamefont {M.~A.}\ \bibnamefont
  {AL-Shudeifat}}, \bibinfo {author} {\bibfnamefont {L.~I.}\ \bibnamefont
  {Manevitch}}, \bibinfo {author} {\bibfnamefont {A.~F.}\ \bibnamefont
  {Vakakis}}, \ and\ \bibinfo {author} {\bibfnamefont {L.~A.}\ \bibnamefont
  {Bergman}},\ }\href@noop {} {\bibfield  {journal} {\bibinfo  {journal}
  {Nonlinear Dynamics}\ }\textbf {\bibinfo {volume} {69}},\ \bibinfo {pages}
  {1693} (\bibinfo {year} {2012})}\BibitemShut {NoStop}%
\bibitem [{\citenamefont {Vitt}\ and\ \citenamefont
  {Gorelik}(1933)}]{Vitt1933}%
  \BibitemOpen
  \bibfield  {author} {\bibinfo {author} {\bibfnamefont {A.}~\bibnamefont
  {Vitt}}\ and\ \bibinfo {author} {\bibfnamefont {G.}~\bibnamefont {Gorelik}},\
  }\href@noop {} {\bibfield  {journal} {\bibinfo  {journal} {Zhurnal
  Tekhnicheskoy Fiziki}\ }\textbf {\bibinfo {volume} {3}},\ \bibinfo {pages}
  {294} (\bibinfo {year} {1933})}\BibitemShut {NoStop}%
\bibitem [{\citenamefont {Kane}\ and\ \citenamefont
  {Kahn}(1968)}]{Kane_JAM1968}%
  \BibitemOpen
  \bibfield  {author} {\bibinfo {author} {\bibfnamefont {T.~R.}\ \bibnamefont
  {Kane}}\ and\ \bibinfo {author} {\bibfnamefont {M.~E.}\ \bibnamefont
  {Kahn}},\ }\href@noop {} {\bibfield  {journal} {\bibinfo  {journal} {Journal
  of Applied Mechanics}\ }\textbf {\bibinfo {volume} {35}},\ \bibinfo {pages}
  {547} (\bibinfo {year} {1968})}\BibitemShut {NoStop}%
\bibitem [{\citenamefont {Tsel'man}(1970)}]{Tselman_JAMM1970}%
  \BibitemOpen
  \bibfield  {author} {\bibinfo {author} {\bibfnamefont {F.~K.}\ \bibnamefont
  {Tsel'man}},\ }\href@noop {} {\bibfield  {journal} {\bibinfo  {journal}
  {Journal of Applied Mathematics and Mechanics}\ }\textbf {\bibinfo {volume}
  {34}},\ \bibinfo {pages} {916} (\bibinfo {year} {1970})}\BibitemShut
  {NoStop}%
\bibitem [{\citenamefont {Rusbridge}(1980)}]{Rusbridge_AmJP1980}%
  \BibitemOpen
  \bibfield  {author} {\bibinfo {author} {\bibfnamefont {M.~G.}\ \bibnamefont
  {Rusbridge}},\ }\href@noop {} {\bibfield  {journal} {\bibinfo  {journal}
  {American Journal of Physics}\ }\textbf {\bibinfo {volume} {48}},\ \bibinfo
  {pages} {146} (\bibinfo {year} {1980})}\BibitemShut {NoStop}%
\bibitem [{\citenamefont {Breitenberger}\ and\ \citenamefont
  {Mueller}(1981)}]{Breitenberger_JMP1981}%
  \BibitemOpen
  \bibfield  {author} {\bibinfo {author} {\bibfnamefont {E.}~\bibnamefont
  {Breitenberger}}\ and\ \bibinfo {author} {\bibfnamefont {R.~D.}\ \bibnamefont
  {Mueller}},\ }\href@noop {} {\bibfield  {journal} {\bibinfo  {journal}
  {Journal of Mathematical Physics}\ }\textbf {\bibinfo {volume} {22}},\
  \bibinfo {pages} {1196} (\bibinfo {year} {1981})}\BibitemShut {NoStop}%
\bibitem [{\citenamefont {Lai}(1984)}]{Lai_AmJP1984}%
  \BibitemOpen
  \bibfield  {author} {\bibinfo {author} {\bibfnamefont {H.~M.}\ \bibnamefont
  {Lai}},\ }\href@noop {} {\bibfield  {journal} {\bibinfo  {journal} {American
  Journal of Physics}\ }\textbf {\bibinfo {volume} {52}},\ \bibinfo {pages}
  {219} (\bibinfo {year} {1984})}\BibitemShut {NoStop}%
\bibitem [{\citenamefont {N{\'{u}}{\~{n}}ez-Y{\'{e}}pez}\ \emph
  {et~al.}(1990)\citenamefont {N{\'{u}}{\~{n}}ez-Y{\'{e}}pez}, \citenamefont
  {Salas-Brito}, \citenamefont {Vargas},\ and\ \citenamefont
  {Vicente}}]{Yepez_PLA1990}%
  \BibitemOpen
  \bibfield  {author} {\bibinfo {author} {\bibfnamefont {H.~N.}\ \bibnamefont
  {N{\'{u}}{\~{n}}ez-Y{\'{e}}pez}}, \bibinfo {author} {\bibfnamefont {A.~L.}\
  \bibnamefont {Salas-Brito}}, \bibinfo {author} {\bibfnamefont {C.~A.}\
  \bibnamefont {Vargas}}, \ and\ \bibinfo {author} {\bibfnamefont
  {L.}~\bibnamefont {Vicente}},\ }\href@noop {} {\bibfield  {journal} {\bibinfo
   {journal} {Physics Letters A}\ }\textbf {\bibinfo {volume} {145}},\ \bibinfo
  {pages} {101} (\bibinfo {year} {1990})}\BibitemShut {NoStop}%
\bibitem [{\citenamefont {Cuerno}, \citenamefont {Ra{\~{n}}ada},\ and\
  \citenamefont {Ruiz-Lorenzo}(1992)}]{Cuerno_AmJP92}%
  \BibitemOpen
  \bibfield  {author} {\bibinfo {author} {\bibfnamefont {R.}~\bibnamefont
  {Cuerno}}, \bibinfo {author} {\bibfnamefont {A.~F.}\ \bibnamefont
  {Ra{\~{n}}ada}}, \ and\ \bibinfo {author} {\bibfnamefont {J.~J.}\
  \bibnamefont {Ruiz-Lorenzo}},\ }\href@noop {} {\bibfield  {journal} {\bibinfo
   {journal} {American Journal of Physics}\ }\textbf {\bibinfo {volume} {60}},\
  \bibinfo {pages} {73} (\bibinfo {year} {1992})}\BibitemShut {NoStop}%
\bibitem [{\citenamefont {Carretero-Gonz{\'{a}}lez}, \citenamefont
  {N{\'{u}}{\~{n}}ez-Y{\'{e}}pez},\ and\ \citenamefont
  {Salas-Brito}(1994)}]{Gonzalez_EJP94}%
  \BibitemOpen
  \bibfield  {author} {\bibinfo {author} {\bibfnamefont {R.}~\bibnamefont
  {Carretero-Gonz{\'{a}}lez}}, \bibinfo {author} {\bibfnamefont {H.~N.}\
  \bibnamefont {N{\'{u}}{\~{n}}ez-Y{\'{e}}pez}}, \ and\ \bibinfo {author}
  {\bibfnamefont {A.~L.}\ \bibnamefont {Salas-Brito}},\ }\href@noop {}
  {\bibfield  {journal} {\bibinfo  {journal} {European Journal of Physics}\
  }\textbf {\bibinfo {volume} {15}},\ \bibinfo {pages} {139} (\bibinfo {year}
  {1994})}\BibitemShut {NoStop}%
\bibitem [{\citenamefont {van~der Weele}\ and\ \citenamefont
  {de~Kleine}(1996)}]{Weele_PhysA1996}%
  \BibitemOpen
  \bibfield  {author} {\bibinfo {author} {\bibfnamefont {J.~P.}\ \bibnamefont
  {van~der Weele}}\ and\ \bibinfo {author} {\bibfnamefont {E.}~\bibnamefont
  {de~Kleine}},\ }\href@noop {} {\bibfield  {journal} {\bibinfo  {journal}
  {Physica A}\ }\textbf {\bibinfo {volume} {228}},\ \bibinfo {pages} {245}
  (\bibinfo {year} {1996})}\BibitemShut {NoStop}%
\bibitem [{\citenamefont {Contopoulos}(1963)}]{Contopoulos_AJ1963}%
  \BibitemOpen
  \bibfield  {author} {\bibinfo {author} {\bibfnamefont {G.}~\bibnamefont
  {Contopoulos}},\ }\href@noop {} {\bibfield  {journal} {\bibinfo  {journal}
  {The Astronomical Journal}\ }\textbf {\bibinfo {volume} {68}},\ \bibinfo
  {pages} {763} (\bibinfo {year} {1963})}\BibitemShut {NoStop}%
\bibitem [{\citenamefont {Hitzl}(1975)}]{Hitzl_CM1975}%
  \BibitemOpen
  \bibfield  {author} {\bibinfo {author} {\bibfnamefont {D.~L.}\ \bibnamefont
  {Hitzl}},\ }\href@noop {} {\bibfield  {journal} {\bibinfo  {journal}
  {Celestial Mechanics}\ }\textbf {\bibinfo {volume} {12}},\ \bibinfo {pages}
  {359} (\bibinfo {year} {1975})}\BibitemShut {NoStop}%
\bibitem [{\citenamefont {Hori}(1966)}]{Hori_ASJ1966}%
  \BibitemOpen
  \bibfield  {author} {\bibinfo {author} {\bibfnamefont {G.-I.}\ \bibnamefont
  {Hori}},\ }\href@noop {} {\bibfield  {journal} {\bibinfo  {journal}
  {Publications of the Astronomical Society of Japan}\ }\textbf {\bibinfo
  {volume} {18}},\ \bibinfo {pages} {287} (\bibinfo {year} {1966})}\BibitemShut
  {NoStop}%
\bibitem [{\citenamefont {Broucke}\ and\ \citenamefont
  {Baxa}(1973)}]{Broucke_CM1973}%
  \BibitemOpen
  \bibfield  {author} {\bibinfo {author} {\bibfnamefont {R.}~\bibnamefont
  {Broucke}}\ and\ \bibinfo {author} {\bibfnamefont {P.~A.}\ \bibnamefont
  {Baxa}},\ }\href@noop {} {\bibfield  {journal} {\bibinfo  {journal}
  {Celestial Mechanics}\ }\textbf {\bibinfo {volume} {8}},\ \bibinfo {pages}
  {261} (\bibinfo {year} {1973})}\BibitemShut {NoStop}%
\bibitem [{\citenamefont {Fermi}\ and\ \citenamefont
  {Rasetti}(1931)}]{Ramaneffekt1931}%
  \BibitemOpen
  \bibfield  {author} {\bibinfo {author} {\bibfnamefont {E.}~\bibnamefont
  {Fermi}}\ and\ \bibinfo {author} {\bibfnamefont {F.}~\bibnamefont
  {Rasetti}},\ }\href@noop {} {\bibfield  {journal} {\bibinfo  {journal}
  {Zeitschrift fur Physik}\ }\textbf {\bibinfo {volume} {71}},\ \bibinfo
  {pages} {689} (\bibinfo {year} {1931})}\BibitemShut {NoStop}%
\bibitem [{\citenamefont {Amat}\ and\ \citenamefont
  {Pimbert}(1965)}]{Amat_JMS1965}%
  \BibitemOpen
  \bibfield  {author} {\bibinfo {author} {\bibfnamefont {G.}~\bibnamefont
  {Amat}}\ and\ \bibinfo {author} {\bibfnamefont {M.}~\bibnamefont {Pimbert}},\
  }\href@noop {} {\bibfield  {journal} {\bibinfo  {journal} {Journal of
  Molecular Spectroscopy}\ }\textbf {\bibinfo {volume} {16}},\ \bibinfo {pages}
  {278} (\bibinfo {year} {1965})}\BibitemShut {NoStop}%
\bibitem [{\citenamefont {Jacob}, \citenamefont {Gross},\ and\ \citenamefont
  {Dreizler}(1978)}]{Jacob_JPB1978}%
  \BibitemOpen
  \bibfield  {author} {\bibinfo {author} {\bibfnamefont {B.}~\bibnamefont
  {Jacob}}, \bibinfo {author} {\bibfnamefont {E.~K.~U.}\ \bibnamefont {Gross}},
  \ and\ \bibinfo {author} {\bibfnamefont {R.~M.}\ \bibnamefont {Dreizler}},\
  }\href@noop {} {\bibfield  {journal} {\bibinfo  {journal} {Journal of Physics
  B}\ }\textbf {\bibinfo {volume} {11}},\ \bibinfo {pages} {3795} (\bibinfo
  {year} {1978})}\BibitemShut {NoStop}%
\bibitem [{\citenamefont {Armstrong}\ \emph {et~al.}(1962)\citenamefont
  {Armstrong}, \citenamefont {Bloembergen}, \citenamefont {Ducuing},\ and\
  \citenamefont {Pershan}}]{Armstrong_PR1962}%
  \BibitemOpen
  \bibfield  {author} {\bibinfo {author} {\bibfnamefont {J.~A.}\ \bibnamefont
  {Armstrong}}, \bibinfo {author} {\bibfnamefont {N.}~\bibnamefont
  {Bloembergen}}, \bibinfo {author} {\bibfnamefont {J.}~\bibnamefont
  {Ducuing}}, \ and\ \bibinfo {author} {\bibfnamefont {P.~S.}\ \bibnamefont
  {Pershan}},\ }\href@noop {} {\bibfield  {journal} {\bibinfo  {journal}
  {Physical Review}\ }\textbf {\bibinfo {volume} {127}},\ \bibinfo {pages}
  {1918} (\bibinfo {year} {1962})}\BibitemShut {NoStop}%
\bibitem [{\citenamefont {Orosco}\ and\ \citenamefont
  {Coimbra}(2016)}]{Orosco2016}%
  \BibitemOpen
  \bibfield  {author} {\bibinfo {author} {\bibfnamefont {J.}~\bibnamefont
  {Orosco}}\ and\ \bibinfo {author} {\bibfnamefont {C.~F.~M.}\ \bibnamefont
  {Coimbra}},\ }\href@noop {} {\bibfield  {journal} {\bibinfo  {journal}
  {Nonlinear Dynamics}\ }\textbf {\bibinfo {volume} {86}},\ \bibinfo {pages}
  {695} (\bibinfo {year} {2016})}\BibitemShut {NoStop}%
\bibitem [{\citenamefont {Orzechowski}\ and\ \citenamefont
  {Fraczek}(2015)}]{Orzechowski2015}%
  \BibitemOpen
  \bibfield  {author} {\bibinfo {author} {\bibfnamefont {G.}~\bibnamefont
  {Orzechowski}}\ and\ \bibinfo {author} {\bibfnamefont {J.}~\bibnamefont
  {Fraczek}},\ }\href@noop {} {\bibfield  {journal} {\bibinfo  {journal}
  {Nonlinear Dynamics}\ }\textbf {\bibinfo {volume} {82}},\ \bibinfo {pages}
  {451} (\bibinfo {year} {2015})}\BibitemShut {NoStop}%
\bibitem [{\citenamefont {Tusset}\ \emph {et~al.}(2016)\citenamefont {Tusset},
  \citenamefont {Piccirillo}, \citenamefont {Bueno}, \citenamefont {Balthazar},
  \citenamefont {Sado}, \citenamefont {Felix},\ and\ \citenamefont
  {Brasil}}]{Tusset2016}%
  \BibitemOpen
  \bibfield  {author} {\bibinfo {author} {\bibfnamefont {A.~M.}\ \bibnamefont
  {Tusset}}, \bibinfo {author} {\bibfnamefont {V.}~\bibnamefont {Piccirillo}},
  \bibinfo {author} {\bibfnamefont {A.~M.}\ \bibnamefont {Bueno}}, \bibinfo
  {author} {\bibfnamefont {J.~M.}\ \bibnamefont {Balthazar}}, \bibinfo {author}
  {\bibfnamefont {D.}~\bibnamefont {Sado}}, \bibinfo {author} {\bibfnamefont
  {J.~L.~P.}\ \bibnamefont {Felix}}, \ and\ \bibinfo {author} {\bibfnamefont
  {R.~M. L. R. d.~F.}\ \bibnamefont {Brasil}},\ }\href@noop {} {\bibfield
  {journal} {\bibinfo  {journal} {Journal of Vibration and Control}\ }\textbf
  {\bibinfo {volume} {22}},\ \bibinfo {pages} {3621} (\bibinfo {year}
  {2016})}\BibitemShut {NoStop}%
\bibitem [{\citenamefont {Nishimura}, \citenamefont {Ikeda},\ and\
  \citenamefont {Harata}(2016)}]{Nishimura2016}%
  \BibitemOpen
  \bibfield  {author} {\bibinfo {author} {\bibfnamefont {K.}~\bibnamefont
  {Nishimura}}, \bibinfo {author} {\bibfnamefont {T.}~\bibnamefont {Ikeda}}, \
  and\ \bibinfo {author} {\bibfnamefont {Y.}~\bibnamefont {Harata}},\
  }\href@noop {} {\bibfield  {journal} {\bibinfo  {journal} {Nonlinear
  Dynamics}\ }\textbf {\bibinfo {volume} {83}},\ \bibinfo {pages} {1705}
  (\bibinfo {year} {2016})}\BibitemShut {NoStop}%
\bibitem [{\citenamefont {Rocha}\ \emph {et~al.}(2017)\citenamefont {Rocha},
  \citenamefont {Balthazar}, \citenamefont {Tusset},\ and\ \citenamefont
  {Piccirillo}}]{Rocha2017}%
  \BibitemOpen
  \bibfield  {author} {\bibinfo {author} {\bibfnamefont {R.~T.}\ \bibnamefont
  {Rocha}}, \bibinfo {author} {\bibfnamefont {J.~M.}\ \bibnamefont
  {Balthazar}}, \bibinfo {author} {\bibfnamefont {A.~M.}\ \bibnamefont
  {Tusset}}, \ and\ \bibinfo {author} {\bibfnamefont {V.}~\bibnamefont
  {Piccirillo}},\ }\href@noop {} {\bibfield  {journal} {\bibinfo  {journal}
  {Journal of Vibration and Control}\ } (\bibinfo {year} {2017})}\BibitemShut
  {NoStop}%
\bibitem [{\citenamefont {Anh}\ \emph {et~al.}(2007)\citenamefont {Anh},
  \citenamefont {Matsuhisa}, \citenamefont {Viet},\ and\ \citenamefont
  {Yasuda}}]{Anh2007}%
  \BibitemOpen
  \bibfield  {author} {\bibinfo {author} {\bibfnamefont {N.~D.}\ \bibnamefont
  {Anh}}, \bibinfo {author} {\bibfnamefont {H.}~\bibnamefont {Matsuhisa}},
  \bibinfo {author} {\bibfnamefont {L.~D.}\ \bibnamefont {Viet}}, \ and\
  \bibinfo {author} {\bibfnamefont {M.}~\bibnamefont {Yasuda}},\ }\href@noop {}
  {\bibfield  {journal} {\bibinfo  {journal} {Journal of Sound and Vibration}\
  }\textbf {\bibinfo {volume} {307}},\ \bibinfo {pages} {187} (\bibinfo {year}
  {2007})}\BibitemShut {NoStop}%
\bibitem [{\citenamefont {Holmes}\ \emph {et~al.}(2006)\citenamefont {Holmes},
  \citenamefont {Full}, \citenamefont {Koditschek},\ and\ \citenamefont
  {Guckenheimer}}]{Holmes2006}%
  \BibitemOpen
  \bibfield  {author} {\bibinfo {author} {\bibfnamefont {P.}~\bibnamefont
  {Holmes}}, \bibinfo {author} {\bibfnamefont {R.~J.}\ \bibnamefont {Full}},
  \bibinfo {author} {\bibfnamefont {D.}~\bibnamefont {Koditschek}}, \ and\
  \bibinfo {author} {\bibfnamefont {J.}~\bibnamefont {Guckenheimer}},\
  }\href@noop {} {\bibfield  {journal} {\bibinfo  {journal} {SIAM Review}\
  }\textbf {\bibinfo {volume} {48}},\ \bibinfo {pages} {207} (\bibinfo {year}
  {2006})}\BibitemShut {NoStop}%
\bibitem [{\citenamefont {Wang}, \citenamefont {Li},\ and\ \citenamefont
  {Xie}(2011)}]{Wang2011}%
  \BibitemOpen
  \bibfield  {author} {\bibinfo {author} {\bibfnamefont {D.}~\bibnamefont
  {Wang}}, \bibinfo {author} {\bibfnamefont {J.}~\bibnamefont {Li}}, \ and\
  \bibinfo {author} {\bibfnamefont {Q.}~\bibnamefont {Xie}},\ }\href@noop {}
  {\bibfield  {journal} {\bibinfo  {journal} {Advances in Structural
  Engineering}\ }\textbf {\bibinfo {volume} {14}},\ \bibinfo {pages} {445}
  (\bibinfo {year} {2011})}\BibitemShut {NoStop}%
\bibitem [{\citenamefont {Castillo-Rivera}\ and\ \citenamefont
  {Tomas-Rodriguez}(2017)}]{Castillo-Rivera2017}%
  \BibitemOpen
  \bibfield  {author} {\bibinfo {author} {\bibfnamefont {S.}~\bibnamefont
  {Castillo-Rivera}}\ and\ \bibinfo {author} {\bibfnamefont {M.}~\bibnamefont
  {Tomas-Rodriguez}},\ }\href@noop {} {\bibfield  {journal} {\bibinfo
  {journal} {Nonlinear Dynamics}\ }\textbf {\bibinfo {volume} {88}},\ \bibinfo
  {pages} {2933} (\bibinfo {year} {2017})}\BibitemShut {NoStop}%
\bibitem [{\citenamefont {Forest}\ and\ \citenamefont
  {Ruth}(1990)}]{Forest1990}%
  \BibitemOpen
  \bibfield  {author} {\bibinfo {author} {\bibfnamefont {E.}~\bibnamefont
  {Forest}}\ and\ \bibinfo {author} {\bibfnamefont {R.~D.}\ \bibnamefont
  {Ruth}},\ }\href@noop {} {\bibfield  {journal} {\bibinfo  {journal} {Physica
  D}\ }\textbf {\bibinfo {volume} {43}},\ \bibinfo {pages} {105} (\bibinfo
  {year} {1990})}\BibitemShut {NoStop}%
\bibitem [{\citenamefont {Tondl}\ \emph {et~al.}(2000)\citenamefont {Tondl},
  \citenamefont {Ruijgrok}, \citenamefont {Verhulst},\ and\ \citenamefont
  {Nabergoj}}]{Tondl_Book2000}%
  \BibitemOpen
  \bibfield  {author} {\bibinfo {author} {\bibfnamefont {A.}~\bibnamefont
  {Tondl}}, \bibinfo {author} {\bibfnamefont {T.}~\bibnamefont {Ruijgrok}},
  \bibinfo {author} {\bibfnamefont {F.}~\bibnamefont {Verhulst}}, \ and\
  \bibinfo {author} {\bibfnamefont {R.}~\bibnamefont {Nabergoj}},\ }\href@noop
  {} {\emph {\bibinfo {title} {{Autoparametric Resonance in Mechanical
  Systems}}}}\ (\bibinfo  {publisher} {Cambridge Univesity Press},\ \bibinfo
  {address} {Cambridge},\ \bibinfo {year} {2000})\BibitemShut {NoStop}%
\bibitem [{\citenamefont {Verhulst}(2002)}]{Verhulst2002}%
  \BibitemOpen
  \bibfield  {author} {\bibinfo {author} {\bibfnamefont {F.}~\bibnamefont
  {Verhulst}},\ }\href@noop {} {\bibfield  {journal} {\bibinfo  {journal} {Acta
  Applicandae Mathematicae}\ }\textbf {\bibinfo {volume} {70}},\ \bibinfo
  {pages} {231} (\bibinfo {year} {2002})}\BibitemShut {NoStop}%
\end{thebibliography}%

\end{document}